\documentclass[prl,twocolumn,showpacs,amssymb]{revtex4}

\usepackage{psfrag,graphicx}
\usepackage{dcolumn}
\usepackage{bm}
\usepackage{amsfonts,amssymb,amsmath}        
\usepackage{yfonts}
\usepackage{xcolor}

\newcommand{\be}{\begin{equation}}
\newcommand{\ee}{\end{equation}}
\newcommand{\bq}{\begin{eqnarray}}
\newcommand{\eq}{\end{eqnarray}}

\newcommand{\rf}[1]{(\ref{#1})}

\begin{document}
\title{Holographic correspondence in topological superconductors} 

\author{Giandomenico Palumbo}
 \affiliation{School of Physics and Astronomy, University of Leeds, Leeds, LS2 9JT, United Kingdom}
\author{Jiannis K. Pachos}
 \affiliation{School of Physics and Astronomy, University of Leeds, Leeds, LS2 9JT, United Kingdom}

\date{\today}

\pacs{03.65.Vf, 11.15.Yc, 11.25.Hf, 73.20.At}

\begin{abstract}

\noindent
We analytically derive a compatible family of effective field theories that uniquely describe topological superconductors in 3D, their 2D boundary and their 1D defect lines. We start by deriving the topological field theory of a 3D topological superconductor in class DIII, which is consistent with its symmetries. Then we identify the effective theory of a 2D topological superconductor in class D living on the gapped boundary of the 3D system. By employing the holographic correspondence we derive the effective chiral conformal field theory that describes the gapless modes living on the defect lines or effective boundary of the class D topological superconductor. We demonstrate that the chiral central charge is given in terms of the 3D winding number of the bulk which by its turn is equal to the Chern number of its gapped boundary.

\end{abstract}

\maketitle

{\bf \em Introduction:--} Topological phases of matter are characterised by topological invariants in the bulk and topological protected gapless edge states \cite{Fradkin, Bernevig}. This bulk-edge correspondence in systems supporting fractional quantum Hall states can be nicely described by the CS$_{2+1}$/CFT$_{1+1}$ correspondence \cite{Witten0}. There, the Chern-Simons (CS) theory that defines the properties of the bulk ground state is in correspondence to the conformal field theory (CFT) that characterises the edge modes. Many of the bulk properties, such as the statistics and fusion of the anyonic quasiparticles, can be derived by simply studying the CFT at the boundary. At the same time, analytically tractable free fermion models, like topological superconductors (TSC) \cite{Ryu}, exhibit fascinating physics. They can support localised Majorana fermions in vortex cores \cite{Jackiw3} with non-Abelian anyonic statistics \cite{Read-Green, Ivanov} that play a central role in topological quantum computation \cite{PachosB}. Moreover, in two dimensions they exhibit gapless Majorana edge modes at their boundary that correspond to CFT with semi-integer central charges, where each Majorana mode contributes 1/2 to their value. Nevertheless, a rigorous holographic correspondence describing the relation between the effective theories of the bulk and the boundary of 2D TSC is still missing.

The goal of this paper is to present a holographic correspondence for TSC of the form TFT$_{3+1}$/TFT$_{2+1}$/CFT$_{1+1}$, where the topological field theories (TFTs) are directly derived from the fermion model of the TSC. We start by considering 3D TSC in the class DIII and show that the corresponding Dirac action is invariant under global SO$(3,2)$ anti-de Sitter (AdS) transformations.
The topological effective theory is then derived by gauging the global symmetry, i.e. by coupling the Dirac fermions to a SO$(3,2)$ Cartan connection \cite{Wise, Randono, Westman}, and then integrating out the fermion fields in the corresponding partition function. At this point, we gap the surface states, e.g. by introducing an external Zeeman field, such that the 2D boundary behaves like a 2D TSC in class D \cite{Finch}.
The corresponding TFT$_{2+1}$ can be derived from the (3+1)-D one by applying the Stokes theorem. This effective theory is formally equivalent to the exotic AdS gravity \cite{Witten} which defines an AdS$_{2+1}$ spacetime. By employing the AdS$_{2+1}$/CFT$_{1+1}$ correspondence \cite{Brown, Kraus, Blagojevic, Klemm}, we can derive an effective CFT with a total chiral central charge $c_{\rm tot}$ which is compatible with 1D Majorana zero modes trapped by defect lines or effective boundaries. These defect lines can be created in several different ways, see e.g. \cite{Qi-Zhang}. Importantly, $c_{\rm tot}$ is given in terms of the winding number, $\nu_\text{3D}$, that characterises the topological phase of the 3D TSC which is equal to the Chern number, $\nu_\text{2D}$, of its gapped boundary \cite{Finch} and its value is not a priori assumed as in \cite{Qi-Zhang}. This implies that $\nu_\text{3D}$ is directly related to the thermal Hall conductance \cite{Cappelli, Stone}, which represents a physical observable associated with the defect lines of the system.

{\bf \em 3D DIII TSC and AdS symmetry:--} We start by considering 3D TSC in the class DIII which preserve time-reversal and particle-hole symmetries. Due to these symmetries the Hamiltonian can be brought in the off-diagonal form \cite{Ryu, Beri} given by
\begin{eqnarray} \label{Hamiltonian}
H =\int d^3x \,\, \psi^{\dagger}(\alpha^{j}p_{j}+i\,\beta \gamma^{5}m)\psi,
\end{eqnarray}
where $j=1,2,3$, $\alpha^{j}=\sigma^{x} \otimes \sigma^{j}$, $\beta =\sigma^{z} \otimes \mathbb{I}_{2 \times 2}$, $\gamma^5 =- i\, \alpha^1\alpha^2\alpha^3$, $\mathbb{I}_{2 \times 2}$ is the identity matrix, and $\sigma^{j}$ are the Pauli matrices. 
The spinor $\psi$ corresponds to a transformed version of a Nambu spinor from the particle-hole basis to the off-diagonal basis \cite{Finch}.
The corresponding action is given by
\begin{eqnarray}
\label{Diracflat}
S^\text{3D}[\psi,\overline{\psi}]=\int d^{4}x\,\,
\overline{\psi}\,(i\,\Gamma^{\mu}\partial_{\mu}-m)\psi,
\end{eqnarray}
where $\mu=0,1,2,3$, $\overline{\psi}=\psi^{\dagger}\Gamma^{0}$, $\Gamma^{0}=i \beta \gamma^{5}$ and $\Gamma^{j}=\Gamma^{0}\alpha^{j}$. This action has a global SO$(3,2)$ AdS symmetry. This symmetry is manifested by the invariance under the transformation
\begin{eqnarray} 
\psi\rightarrow  e^{\frac{i}{2}\,\theta_{AB} J^{AB}}\psi\hspace{0.3cm}\,\,\text{and}\,\,
\psi^{\dagger}\rightarrow \psi^{\dagger} e^{-\frac{i}{2}\,\theta_{AB} J^{AB}},
\label{eqn:trans}
\end{eqnarray}
with $A, B=0,1,2,3,4$, where $\theta_{AB}$ is an anti-symmetric parameter and $J^{AB}=\left(J^{AB}\right)^{\dagger}$ are the generators of the AdS algebra. These generators satisfy the commutation relations 
\begin{align}\label{symmetry}
& \left[J^{AB},J^{CD}\right]= \hspace{2.3cm}\nonumber \\ 
& -i\,\left(\eta^{AC}J^{BD}-\eta^{AD}J^{BC}-\eta^{BC}J^{AD}+\eta^{BD}J^{AC}\right),
\end{align}
where $\eta^{AB}={\rm diag}(-1,+1,+1,+1,-1)$. Let us consider the matrices $\Gamma^{A}=(\Gamma^\mu, \gamma^{5})$, satisfying the Clifford algebra $\{\Gamma^{A},\Gamma^{B}\}=-2\,\eta^{AB}\mathbb{I}_{4\times 4}$. The generators $J^{AB}=-i\,[\Gamma^{A},\Gamma^{B}]/4$ belong to the Clifford representation of Spin$(3,2)$, which is the double covering of SO$(3,2)$ \cite{Randono, Westman}. With this representation one can show that the transformations (\ref{eqn:trans}) are the only non-trivial orthogonal ones that leave invariant the Dirac mass in action (\ref{Diracflat}). Moreover, $\Gamma^{\mu}$ in the kinetic term of the action transforms as a vector under SO$(3,2)$ so that $\overline\psi i \Gamma^{\mu} \partial_{\mu} \psi$ is a scalar under \rf{eqn:trans}. Note also that due to the AdS invariance, the adjoint spinor $\overline{\psi}$ assumes a different form than the more familiar Lorentz-invariant one (i.e. $\overline{\psi}=\psi^{\dagger}\beta$) \cite{Westman}.

In order to derive the topological effective theory that describes the 3D DIII TSC in the low energy regime, we first gauge the global SO$(3,2)$ symmetry of action \rf{Diracflat}. In other words, we introduce the tetrads $e_{A}^{\mu}$ and replace the standard derivative with a covariant derivative $D_{\mu}=\partial_{\mu}+A_{\mu}$, where $A_{\mu}=(i/2)A^{AB}_{\mu}J_{AB}$ is the connection that takes values in the AdS algebra \cite{Randono, Westman}. On a Lorentzian curved spacetime, the corresponding gauged action is given by 
\begin{eqnarray}\label{Dirac2}
S^\text{3D}[\psi,\overline{\psi}, A_{\mu}]=\int d^{4}x\,|e|\,
\overline{\psi}\,(i\,\widehat{\Gamma}^{\mu}D_{\mu}-m)\psi,
\end{eqnarray}
where $|e|$ is the determinant of $e_{A}^{\mu}$, $\widehat{\Gamma}^{\mu}=e_{A}^{\mu}\Gamma^{A}$ with $\Gamma^{4}=\gamma^{5}$ \cite{Westman}. Clearly, in the flat limit, 
$|e|=1$, $A_{\mu}=0$ and $e_{A}^{\mu}\Gamma^{A}=\delta_{A}^{\mu}\Gamma^{A}=\Gamma^{\mu}$, we recover (\ref{Diracflat}). 

To derive the above action we have used a generalisation of Riemannian geometry called Cartan geometry \cite{Wise, Randono}, with $A_{\mu}$ called Cartan connection, where the tangent space is isomorphic to the AdS space, a manifold with a globally constant {\em negative curvature}. 
It has been shown that the AdS tangent space is compatible with the existence of Majorana fermions \cite{Chamseddine}. Moreover, in this geometric framework, spin connection and tetrads are independent variables, implying a non-zero torsion at the level of field action \cite{Hehl}, as we shall see in the following. 

{\bf \em Effective topological field theory:--} To find the TFT$_{3+1}$ that corresponds to the 3D TSC we have to integrate out the fermion field in the partition function of $S^\text{3D}[\psi,\overline{\psi}, A_{\mu}]$. The resulting effective action $S^\text{3D}_{\rm eff}[A_{\mu}]$ defined by 
\begin{eqnarray}
e^{i S^\text{3D}_{\rm eff}[A_{\mu}]}=\int D\overline{\psi}\,\textit{D}\psi\,e^{i S^\text{3D}[\psi,\overline{\psi}, A_{\mu}]},
\end{eqnarray}
can be divided in a topological and a non-topological part. The topological part, $S_{\rm top}^\text{3D}[A_{\mu}]$, is dominant at low energies and describes the large distance physics of the ground state. In general, even if the above path integral is bilinear in the fermionic operators, it is not possible to calculate precisely the non-topological part of the effective action. On the other hand, $S_{\rm top}^\text{3D}[A_{\mu}]$ can be derived exactly. In fact, it is possible to show that after a Wick rotation, the corresponding Euclidean version of the topological action is proportional to the analytic index of the Dirac operator $\widehat{\Gamma}^{\mu}D_{\mu}$ \cite{Palumbo-Index}. Thus, the Lorentzian action $S_{\rm top}^\text{3D}[A_{\mu}]$ is proportional to a topological invariant \cite{Eguchi} and the coefficient of proportionality can be fixed by physical motivations as shown in \cite{Wang-Zhang}. As a result we have
\begin{eqnarray}
S_{\rm top}^\text{3D}[A_{\mu}]=k\,\int d^{4}x\,
\epsilon^{\mu\nu\alpha\beta}\, \text{tr}\, F_{\mu\nu}F_{\alpha\beta},
\end{eqnarray}
with 
\be
k={\nu_\text{3D}\over192 \pi}, 
\label{eqn:k}
\ee
where $\nu_\text{3D}$ is the 3D winding number, i.e. the topological index that characterise the bulk of 3D TSC \cite{Ryu}. Moreover, $F_{\mu\nu}=[D_{\mu}, D_{\nu}]$ is the curvature tensor of the Cartan connection $A_\mu$, and the trace is taken over the gauge index. 

Similarly to Riemannian geometry, in Cartan geometry, the Cartan connection gives a prescription for parallel transport from one tangent space to another, along a path defined on a curved spacetime. These transports can be decomposed in moves that do not change the point of contact (spinning around the point of contact without rolling) and moves that change the point of contact (rolling without slipping) \cite{Wise, Randono}. At algebraic level, this means that the Lie algebra $g=so(3,2)$ can decompose into the Lorentz sub-algebra $h=so(3,1)\subset so(3,2)$ and its complement $p=so(3,2)/so(3,1)$, i.e. there exists a Killing-orthogonal splitting such that \cite{Wise, Randono}
\begin{eqnarray}
g=h\oplus p.
\end{eqnarray}
This implies that the connection $A_{\mu}$ can be written in terms of spin connection $\omega_{\mu}^{ab}$ and the tetrads $e_{\mu}^a$ ($a,b=0,1,2,3$) where the former is related to the local SO$(3,1)$ Lorentz transformations and the latter to the local spacetime translations. Explicitly, we have that
\begin{eqnarray}\label{Cartan}
A_{\mu}=\omega_{\mu}+\frac{1}{l}\,e_{\mu}=\frac{i}{4}[\gamma_{a},\gamma_{b}]\,\omega_{\mu}^{ab}+\frac{i}{2\,l}\,\gamma_{a}\,e_{\mu}^{a},
\end{eqnarray}
where $l$ is a dimensionful real parameter as the tetrads are dimensionless. The specific value of $l$ is related to curvature radius of the AdS tangent space and thus not relevant here as we are only concerned with the topological characteristics of the AdS space. In \rf{Cartan} the $4 \times 4$ Dirac matrices $\gamma_{a}= (i\gamma^{5}\Gamma_{0}, i\gamma^{5}\Gamma_{j})$ are related to the Clifford algebra representation of Spin$(3,1)$, namely the double covering of SO$(3,1)$ and satisfy the anti-commutation condition $\{\gamma_{a},\gamma_{b}\}=-2\,\eta_{ab}\,\mathbb{I}_{4\times 4}$. In this way, the corresponding curvature tensor $F_{\mu\nu}$ is given by
\begin{eqnarray}
F_{\mu\nu}=R_{\mu\nu}-\frac{1}{4\,l^{2}}(e_{\mu}^{a}e_{\nu}^{b}-e_{\nu}^{a}e_{\mu}^{b})[\gamma_{a},\gamma_{b}]+\frac{1}{l}\,T_{\mu\nu},
\end{eqnarray}
with $R_{\mu\nu}=R_{\mu\nu}^{ab}\,i\,[\gamma_{a},\gamma_{b}]/4$ and $T_{\mu\nu}=\frac{i}{2}\gamma_{a}T^{a}_{\mu\nu}$, where 
\begin{align}
R_{\mu\nu}^{ab} & =\partial_{\mu} \omega_{\nu}^{ab}-\partial_{\nu}\omega_{\mu}^{ab}+\omega_{\mu\,\,c}^{a}\omega_{\nu}^{cb}-\omega_{\nu\,\,c}^{a}\omega_{\mu}^{cb}, \nonumber \\
T^{a}_{\mu\nu} & =\partial_{\mu}e_{\nu}^{a}-\partial_{\nu}e_{\mu}^{a}+\omega_{\mu\,\, b}^{a}e_{\nu}^{b}-\omega_{\nu\,\, b}^{a}e_{\mu}^{b}, \hspace{0.5cm}
\end{align}
represent the Riemann and the torsion tensors, respectively.
We can now rewrite the topological action as follows \cite{Zanelli}
\begin{align}
& S_\text{top}^\text{3D}[\omega_{\mu},e_{\mu}]= \nonumber \\
& k\,\int d^{4}x\,
\epsilon^{\mu\nu\alpha\beta}\, \text{tr}\, \Big[ R_{\mu\nu}R_{\alpha\beta}+ 
{1\over l^{2}}\Big(T_{\mu\nu}T_{\alpha\beta}-R_{\mu\nu}e_{\alpha}e_{\beta}\Big)\Big].
\end{align}
Note that the first term is proportional to the Pontryagin invariant \cite{Eguchi} while the second term is proportional to the Nieh-Yan topological term \cite{Nieh}. 

In order to have a 2D topological phase on the boundary of the 3D system, we initially consider a 3D bulk topological equivalent to a three-torus and introduce the boundary created by breaking the periodicity in one of the spacial dimensions. In this way, there appear two disconnected surfaces that support gapless helical Majorana modes. We can gap them, e.g. by introducing a Zeeman field, such that the gapped boundary comprising of {\em both} surfaces behaves like a 2D TSC in the class D \cite{Finch}. Note that we choose a suitable time-reversal-breaking external field such that the Majorana gapped modes share the same mass sign. Then a Z winding number can be defined that classifies the topological phase of the boundary. This Z number is in agreement with the 3D winding number that, by its turn, classifies the 3D TSC. In this way, also when the 3D winding number is even, its $Z_2$ ambiguity is avoided \cite{Qi-Zhang, Finch}.
 As $S_{\rm top}^\text{3D}$ is a total derivative we can employ the Stokes theorem. The corresponding topological effective action $S_{\rm top}^\text{2D}$ on each disconnected surface is given by
\begin{align}
\label{DoubleCS}
& S_{\rm top}^\text{2D}[\omega_{\mu},e_{\mu}] = \nonumber \\  
& k\,\int d^{3}x\, \epsilon^{\mu\nu\lambda}
 \text{tr} \left(\omega_{\mu}\partial_{\nu}\omega_{\lambda}+\frac{2}{3}\,\omega_{\mu}\omega_{\nu}\omega_{\lambda}+\right.  \left. \frac{1}{l^{2}}T_{\mu\nu}e_{\lambda}\right)=\nonumber \\ 
 & \frac{k}{2}\left(\text{CS}[A^{+}_{\mu}]+\text{CS}[A^{-}_{\mu}]\right),
\end{align}
i.e. it is equivalent to a double CS theory, with
\begin{eqnarray}
\text{CS}[A^{\pm}_{\mu}]=\int d^{3}x\,\epsilon^{\mu\nu\lambda}
 \text{tr} \Big(A^{\pm}_{\mu}\partial_{\nu}A^{\pm}_{\lambda}+\frac{2}{3}A^{\pm}_{\mu}A^{\pm}_{\nu}A^{\pm}_{\lambda}\Big).
\end{eqnarray}
Each Cartan connection $A^{\pm}_{\mu}=\omega_{\mu}\pm \frac{1}{l}\,e_{\mu}$ ($\mu=0,1,2$) takes values in the (2+1)-D Lorentz algebra $so(2,1)$ \cite{Wise2}, such that $so(2,1)\times so(2,1)\simeq so(2,2)$, where $so(2,2)$ is the (2+1)-D AdS algebra. In this way, the above action describes the exotic AdS gravity \cite{Witten}, which is different from the standard Einstein-Hilbert action. In the latter, the torsion tensor does not appear while the parity invariance is preserved because the Einstein theory can be written as a {\em difference} between two CS theories.

Interestingly, in a purely 2D TSC Hamiltonian in class D, the off-diagonal basis does not exist due to the lack of the chiral symmetry \cite{Ryu}. This has important consequence for the gauging procedure if we were to start from this 2D system. In that case the corresponding (2+1)-D Dirac action with a Dirac mass, is invariant with respect to both global SO$(2,1)$ and SO$(2,2)$ transformations. This is due to the fact that SO$(2,2)$ can be seen as a double copy of the SO$(2,1)$ Lorentz group as shown above for the corresponding algebras. Thus, the choice of the (2+1)-D AdS group is not unique as it is in the 3D case. 

{\bf \em Lorentz anomaly and chiral central charge:--} We now employ the AdS$_{2+1}$/CFT$_{1+1}$ correspondence \cite{Brown, Kraus, Blagojevic, Klemm} to calculate the total chiral central charge, $c_{\rm tot}$, corresponding to the boundary of the 3D TSC. The importance of the central charge is in determining the physics of the gapless modes emerging at the defect lines of the system. We now show that the $c_{\rm tot}$ of TSC indeed corresponds to Majorana fermions.

One of the main characteristics of AdS$_{2+1}$ gravity is the existence of a holographic stress-energy tensor \cite{Kraus}, defined on the {\em asymptotic boundary} of AdS$_{2+1}$. 
Here, we assume that the (2+1)-D manifold is asymptotically diffeomorphic to $\Sigma_{2}\times \mathbb{R}$, where $\Sigma_{2}$ is the spacetime where the dual CFT resides. The local coordinates are denoted by $x^{\mu}=(x^{i}, \rho)$, with $i=0,1$ while $\rho$ is the radial coordinate. Note that because $\Sigma_2$ has zero torsion \cite{Klemm}, the (1+1)-D spin connection and the corresponding Riemann tensor $R_{ij}^{uv}$ ($u,v=0,1$) can be written directly in terms of the zweibein $e_{u}^{i}$ associated to the geometry of $\Sigma_2$.

The holographic stress-energy tensor, $\tau^{u}_{i}$, is defined in the limit $\rho\rightarrow \infty$ by varying the exotic gravity action $S_{\rm top}^\text{2D}$ on $\Sigma_{2}$ with respect to $e_{u}^{i}$ \cite{Klemm}
\begin{eqnarray}
\tau^{u}_{i}=\frac{2\pi}{|e|} \left. \frac{\delta S_{\rm top}^\text{2D}}{\delta e^{i}_{u}}\right|_{\rho\rightarrow \infty}.
\label{eqn:torsion11}
\end{eqnarray}
Importantly, this boundary tensor, $\tau^{uv}=\tau^{u}_{i}e^{i v}$, is not symmetric giving rise to the antisymmetric part, $\tau^{[uv]}=\tau^{uv}-\tau^{vu}$. From \rf{eqn:torsion11} we find that 
\begin{eqnarray}\label{Lorentz1}
\tau^{[uv]} = \frac{\pi\,k}{|e|}\epsilon^{ij}\,R_{ij}^{uv}.
\end{eqnarray}
The failure of the stress-energy tensor to be symmetric demonstrates that 
on the (1+1)-D asymptotic boundary there is a Lorentz anomaly \cite{Jackiw}. This quantum anomaly appears in chiral CFT and is proportional to the chiral central charge c \cite{Cappelli, Stone}
\begin{eqnarray}\label{Lorentz}
\tau_\text{CFT}^{[uv]} = \frac{c}{48 |e|}\,\epsilon^{ij}\,R_{ij}^{uv}.
\end{eqnarray}
Due to the AdS$_{2+1}$/CFT$_{1+1}$ correspondence, relations (\ref{Lorentz1}, \ref{Lorentz}) imply that $c=48\pi\,k$ \cite{Blagojevic, Klemm}. This reflects the fact that the exotic AdS gravity breaks parity and time-reversal symmetry, being equivalent to a {\em sum} of two CS theories, as shown in (\ref{DoubleCS}).

For a finite 3D TSC in class DIII bounded by two gapped disconnected surfaces there exists a correspondence between the 3D winding number, $\nu_\text{3D}$, and the 2D Chern number, $\nu_\text{2D}$, describing the 2D TSC in class D of its boundary consisting of {\em both} surfaces \cite{Finch}. This implies that in order to calculate the total chiral central charge $c_{\rm tot}$ we have to sum the two chiral central charges corresponding to the two gapped surfaces each one described by an AdS gravity as derived above. Taking into account \rf{eqn:k} we have that
\begin{eqnarray}
\label{central}
c_{\rm tot}=2\, c=\frac{1}{2}\,\nu_\text{3D}=\frac{1}{2}\,\nu_\text{2D}.
\end{eqnarray}
This is the central result of our work. It signifies that the CFT nested at 1D defect lines has degrees of freedom determined by the winding number of the TSC. As each Majorana edge mode contributes 1/2 to the central charge, we deduce that each defect line introduced on the surface of our 3D TSC can be described by a chiral CFT that supports $N=\nu_\text{2D}=\nu_\text{3D}$ Majorana zero-modes, in agreement with the bulk boundary correspondence of TSC \cite{Bernevig, Ryu, Beri}. 

{\bf \em Winding numbers and thermal currents:--} 
It is well-known that the existence of a Chern number in the bulk of quantum Hall states reflects the presence of a quantised and experimentally measurable charged Hall conductance \cite{Thouless}. This clear picture is unsatisfactory in the case of 3D systems described by a 3D winding number where there is not an equivalent bulk transport property. Moreover, due to the lack of charge conservation in TSC we do not have quantised charged currents, but only {\em quantised thermal currents}. Nevertheless, these currents cannot exist in the 3D bulk of TSC nor at their gapped 2D boundary \cite{Stone}. In these systems, the thermal Hall conductance $\kappa_{\rm th}$ has been already analysed from a geometric prospective in \cite{Read-Green, Qi-Zhang, Ryu-Moore} by connecting it to the gravitational SO$(2,1)$ Chern-Simons theory \cite{Jackiw2}. \\
However, as observed in \cite{Stone}, $\kappa_{\rm th}$ should be confined to one dimensional defect lines (or edge states), and cannot flow in the 2D bulk. In fact, in critical 1D systems described by a chiral CFT, there exists a well-defined $\kappa_{\rm th}$ proportional to $c$, given by
\begin{eqnarray}\label{ThermalHall}
\kappa_{\rm th}=\frac{\pi}{6}\,c\,T,
\end{eqnarray}
where $T\rightarrow 0$ is the temperature and $\hbar=k_{\rm b}=1$, where $k_{\rm b}$ is the Boltzmann constant.
This quantum effect is related to the presence of a (1+1)-D gravitational (Lorentz) anomaly \cite{Cappelli, Stone}. Hence, in our case, the physical interpretation of the 3D winding number, $\nu_\text{3D}$, of TSC needs to be addressed. 

Our derivation of a compatible family TFT$_{3+1}$/AdS$_{2+1}$/CFT$_{1+1}$ of effective field theories gives a physical interpretation to $\nu_\text{3D}$ and $\nu_\text{2D}$ in terms of \rf{central}. The emerging gapless phases live along the defect lines created on the gapped boundary of 3D system. Our relation between the winding numbers of the TFTs and the central charge of the CFT establishes the interpretation of $\nu_\text{2D}$ and $\nu_\text{3D}$ in terms of the induced 1D thermal Hall conductance (\ref{ThermalHall}) with chiral central charge $c_{\rm tot}$. 

{\bf \em Conclusions:--} In this letter we have derived the compatible family of effective descriptions of TSC in 3D, 2D and 1D. This consistent approach allowed us to evaluate the chiral central charge $c_{\rm tot}$ associated with the thermal current of 3D TSC directly from its 3D winding number rather than assuming a priori its value \cite{Qi-Zhang}. Our approach is based on the gauging procedure of the symmetries present in the 3D system. This process uniquely gives rise to an effective curved spacetime which is described by the Cartan geometry. This geometry represents the natural framework of the exotic AdS gravity because of the presence of a non-zero torsion tensor. 
Note that this tensor has been employed in the geometric description of lattice deformations and Hall viscosity \cite{Zaanen, Cortijo, Hughes, Hughes1} and in the context of fractional quantum Hall effect \cite{Abanov, Son, Read}. Our work demonstrates that the torsion tensor is naturally introduced by gauging the symmetries present in the system.
Thus, the AdS$_{2+1}$/CFT$_{1+1}$ holographic correspondence can be directly employed for the study of the defect lines and their associated physical observables. 
Note that even if the holographic correspondence has been already applied in condensed matter as a survey tool in a more abstract way \cite{Hartnoll, McGreevy, Sachdev, Green}, in our case it has been motivated by the properties of the microscopic model.

{\bf \em Acknowledgments:--} We thank Duncan Haldane, Roman Jackiw and Michael Stone for inspiring conversations. This work was supported by EPSRC.


\begin{thebibliography}{25}


\bibitem{Fradkin}
E. Fradkin, \emph{Field Theories of Condensed Matter Physics}, Cambridge University Press (2013).

\bibitem{Bernevig}
B. A. Bernevig, T. L. Hughes, \emph{Topological Insulators and Topological Superconductors}, Princeton University Press (2013).

\bibitem{Witten0}
E. Witten, Comm. Math. Phys. \textbf{121}, 351 (1989).

\bibitem{Ryu} 
S. Ryu, A. P. Schnyder, A. Furusaki and A. W. W. Ludwig, New J. Phys. \textbf{12}, 065010 (2010).

\bibitem{Jackiw3}
R. Jackiw and P. Rossi, Nucl. Phys. B \textbf{190}, 681 (1981).

\bibitem{Read-Green}
N. Read and D. Green, Phys. Rev. B \textbf{61}, 10267 (2000).

\bibitem{Ivanov}
D. A. Ivanov, Phys. Rev. Lett. \textbf{86}, 268 (2001).


\bibitem{PachosB}
J. K. Pachos, \emph{Topological Quantum Computation}, Cambridge University Press (2012).

\bibitem{Wise}
D. K. Wise, Class. Quantum Grav. \textbf{27}, 155010 (2010).

\bibitem{Randono}
A. Randono, arXiv:1010.5822.


\bibitem{Westman}
H. F. Westman and T. G. Zlosnik, Ann. Phys. \textbf{334}, 157 (2013).

\bibitem{Finch}
P. Finch, J. de Lisle, G. Palumbo, and J. K. Pachos, Phys. Rev. Lett. \textbf{114}, 016801 (2015).

\bibitem{Witten}
E. Witten, Nucl. Phys. B \textbf{311}, 46 (1988).


\bibitem{Brown}
E. J. D. Brown and M. Henneaux, Comm. Math. Phys. \textbf{104}, 207 (1986).

\bibitem{Kraus}
V. Balasubramanian and P. Kraus, Comm. Math. Phys. \textbf{208}, 413 (1999).


\bibitem{Blagojevic}
M. Blagojevic and M. Vasilic, Phys. Rev. D \textbf{68}, 104023 (2003).


\bibitem{Klemm}
D. Klemm and G. Tagliabue, Class. Quantum Grav. \textbf{25}, 035011 (2008).


\bibitem{Qi-Zhang}
Z. Wang, X.-L. Qi and S.-C. Zhang, Phys. Rev. B \textbf{84}, 014527 (2011).

\bibitem{Cappelli}
A. Cappelli, M. Huerta and G. R. Zemba, Nucl. Phys. B
\textbf{636}, 568 (2002).

\bibitem{Stone}
M. Stone, Phys. Rev. B \textbf{85}, 184503 (2012).

\bibitem{Beri} 
B. B\' eri, Phys. Rev. B \textbf{81}, 134515 (2010).

\bibitem{Chamseddine}
A. H. Chamseddine and V. Mukhanov, JHEP \textbf{1311}, 95 (2013).


\bibitem{Hehl}
F. W. Hehl, P. von der Heyde, G. D. Kerlick and J. M. Nester,
Rev. Mod. Phys. \textbf{48}, 393 (1976).






\bibitem{Palumbo-Index} 
G. Palumbo, R. Catenacci and A. Marzuoli, Int. J. Mod. Phys. B \textbf{28}, 1350193 (2014).

\bibitem{Eguchi} 
T. Eguchi, P. B. Gilkey and A. J. Hanson, Phys. Rep.
\textbf{66}, 213 (1980).


\bibitem{Wang-Zhang}
Z. Wang and S.-C. Zhang, Phys. Rev. B \textbf{86}, 165116 (2012).


\bibitem{Zanelli}
O. Chandía and J. Zanelli, Phys. Rev. D \textbf{55}, 7580 (1997).

\bibitem{Nieh}
H. T. Nieh and M. L. Yan, J. Math. Phys. \textbf{23}, 373 (1982).

\bibitem{Wise2}
D. K. Wise, SIGMA \textbf{5}, 080 (2009).




\bibitem{Jackiw}
S. Treiman, R. Jackiw, B. Zumino and E. Witten, \emph{Current Algebra and Anomalies}, World Scientific (1985).

\bibitem{Thouless}
D. J. Thouless, M. Kohmoto, M. P. Nightingale and M. den Nijs,
Phys. Rev. Lett. \textbf{49}, 405 (1982).

\bibitem{Ryu-Moore}
S. Ryu, J. E. Moore and A. W. W. Ludwig, Phys. Rev. B \textbf{85}, 045104 (2012).

\bibitem{Jackiw2}
S. Deser, R. Jackiw, and S. Templeton, Phys. Rev. Lett. \textbf{48}, 975 (1982).

\bibitem{Zaanen}
A. Mesaros, D. Sadri and J. Zaanen,
Phys. Rev. B \textbf{82}, (2010).


\bibitem{Cortijo}
F. de Juan, A. Cortijo and M. A.H. Vozmediano, Nucl. Phys. B
\textbf{828}, 625 (2010).

\bibitem{Hughes}
T. L. Hughes, R. G. Leigh and E. Fradkin
Phys. Rev. Lett. \textbf{107}, 075502 (2011).

\bibitem{Hughes1}
T. L. Hughes, R. G. Leigh and O. Parrikar
Phys. Rev. D \textbf{88}, 025040 (2013).

\bibitem{Abanov}
A. Gromov and A. G. Abanov,
Phys. Rev. Lett. \textbf{114}, 016802 (2015).

\bibitem{Son}
M. Geracie, D. T. Son, C. Wu and S.-F. Wu,
Phys. Rev. D \textbf{91}, 045030 (2015).

\bibitem{Read}
B. Bradlyn and N. Read,
Phys. Rev. B \textbf{91}, 125303 (2015).


\bibitem{Hartnoll}
S. A. Hartnoll, Class. Quant. Grav. \textbf{26}, 224002 (2009).

\bibitem{McGreevy}
J. McGreevy, Adv. High Energy Phys. \textbf{2010}, 723105 (2010).

\bibitem{Sachdev}
S. Sachdev, Ann. Rev. Cond. Matt. Phys. \textbf{3}, 9 (2012).

\bibitem{Green}
A. G. Green, Contemporary Physics \textbf{54}, 33 (2013).


\end{thebibliography}
\end{document}